# Electric Field Control of Magnetic Skyrmion Helicity in a Centrosymmetric 2D van der Waals Magnet

Myung-Geun Han[1§*], Joachim Dahl Thomsen[2,3§], John P. Philbin[3], Junsik Mun[1], Eugene Park[2], Fernando Camino[4], Lukáš Děkanovský[5], Chuhang Liu[1,6], Zdenek Sofer[5], Prineha Narang[3], Frances M. Ross[2] and Yimei Zhu[1]

[1]Condensed Matter Physics and Materials Science, Brookhaven National Laboratory, Upton, New York 11973, USA

[2]Department of Materials Science and Engineering, Massachusetts Institute of Technology, Cambridge, Massachusetts 02139, USA

[3]Division of Physical Sciences, College of Letters and Science, University of California, Los Angeles, California 90095, USA

[4]Center for Functional Nanomaterials, Brookhaven National Laboratory, Upton, New York 11973, USA

[5]Department of Inorganic Chemistry, University of Chemistry and Technology Prague, Technická 5, Prague 6, 166 28, Czech Republic

[6]Department of Physics and Astronomy, Stony Brook University, Stony Brook, New York 11794, USA

These authors contributed equally: Myung-Geun Han, Joachim Dahl Thomsen

[*]Email: mghan@bnl.gov

## Abstract

Two-dimensional van der Waals magnets hosting topological magnetic textures, such as skyrmions, show promise for spintronics and quantum computing. Electrical control of these topological spin textures is crucial for enhancing operational performance and functionality. Here, using electron microscopy combined with *in situ* electric and magnetic biasing, we show that the skyrmion helicity, whether left-handed or right-handed, in insulating $Cr_2Ge_2Te_6$, can be controlled by the direction of external electric field applied during the magnetic field cooling process. The electric field-tuned helicity remains stable, even amid variations in magnetic and electric fields. Our theoretical investigation reveals that nonzero Dzyaloshinskii-Moriya interactions between nearest neighbors, induced by the electric field, change their sign upon reversing the electric field direction, thereby facilitating helicity selection. The electrical control of magnetic helicity demonstrated in this study can be extended to other centrosymmetric skyrmion-hosting magnets, paving the way for future device designs in topological spintronics and quantum computing.

Magnetic skyrmions and topologically nontrivial spin structures in two-dimensional (2D) van der Waals (vdW) materials have been studied extensively due to their relevance fundamental spin physics in low dimensions and their potential spintronic applications.[1-3] These topologically protected spin structures exhibit particle-like stability, making them promising information carriers in spintronic devices[4]. In 2D vdW materials, the integration of skyrmions with other 2D vdW layers featuring diverse and tunable physical properties offers exciting opportunities for designing novel states with emergent properties and behaviors.[5-7] Experimentally, topological spin structures have been reported in various 2D vdW ferromagnetic materials, including skyrmionic bubbles in $Cr_2Ge_2Te_6$ (CGT)[8] with perpendicular uniaxial anisotropy, and Néel-type skyrmions in oxidized $Fe_3GeTe_2$[9] and $WTe_2/Fe_3GeTe_2$.[10] There are two primary types of skyrmions in ferromagnetic 2D vdW magnets, based on their stabilization mechanisms: one with Dzyaloshinskii-Moriya interaction (DMI) and the other without DMI. Analogous to bulk chiral magnets,[11] DMI in non-centrosymmetric crystal structures stabilizes chiral spin arrangements.[12] The chirality of DMI-stabilized skyrmions is determined by the crystal chirality. Conversely, in centrosymmetric crystals with uniaxial anisotropy, such as CGT,[8] dipolar interactions stabilize Bloch-type skyrmionic bubbles, which are topologically equivalent to Bloch-type skyrmions in chiral magnets. However, due to the absence of DMIs, the helicity of skyrmionic bubbles in centrosymmetric magnets is random. Here, the helicity represents the in-plane rotation sense, either clockwise (CW) or counterclockwise (CCW).

The ability to control the helicity of skyrmionic spin structures by electric means is crucial for advancing spintronic applications and designing novel states in 2D heterostructures[13-15] and quantum computing applications.[16] In particular, deterministic control of skyrmion helicity can provide a viable method to realize skyrmion-based quantum qubit where electric field control of skyrmion helicity is required[16].

We note that electric fields can polarize insulating ferromagnetic 2D vdW materials. In a polarized state, nonzero DMIs may arise as the centrosymmetry is broken. Theoretical studies have predicted that out-of-plane electric fields can induce skyrmionic spin structures in insulating 2D vdW $CrI_3$ (~ 1 eV bandgap[17]) monolayer[18-19] and CGT (~ 0.4 eV bandgap[20]) heterostructures.[21] As the sign of electric-field-induced DMIs is switchable by reversing the direction of the electric field, the magnetic helicity in skyrmionic spin structures can be controlled in insulating or semiconducting 2D ferromagnets.[22] Additionally, the magnetic properties of CGT can be manipulated with electric stimuli. For instance, electrogating experiments performed on CGT layers with an ionic gel showed not only a dramatic increase in $T_C$ (from 60 K to 200 K), but also an easy axis rotation from out-of-plane to in-plane direction, significantly altering magnetic spin structures.[23] A recent study demonstrated that the spin helicity in a metallic helimagnet MnP can be controlled by spin-transfer torque when the current direction is parallel to the external magnetic field applied.[24] Notably, two studies have demonstrated the electric field control of Néel-type skyrmion chiralities in multilayered magnetic films.[25-26] Nevertheless, there has been no experimental work on the control of topological magnetic spin structures by external electric fields in insulating or semiconducting 2D vdW magnets.

In this letter, we integrate *in situ* electric and magnetic fields with cryogenic Lorentz transmission electron microscopy (LTEM) to experimentally demonstrate the tunable helicity of skyrmionic bubbles in CGT flakes. Our data reveal that the helicity (the direction of in-plane spin rotation) of Bloch-type skyrmionic bubbles can be reversed by changing the electric field direction, without altering the polarity (the core spin direction), thus favoring one of the two possible helicities by an external electric field. Our density functional theory (DFT) calculations support these experimental observations, attributing the helicity selection in skyrmionic bubbles to the electric-field-induced DMIs between nearest neighbors. Achieving helicity control via an electric field, as demonstrated in this study, opens up promising avenues for designing novel spintronic devices and quantum qubits based on topological spin structures in the 2D vdW heterostructures.

Figure 1a shows a few-layer graphene (FLG)/CGT/FLG heterostructure fabricated over an electron-beam imaging window (vacuum), patterned in a custom designed silicon nitride membrane chip with two pre-patterned electrodes (Norcada Inc., Fig. S1 in the Supplementary Information). Due to the bandgap of CGT (~0.4 eV), no significant current is expected to flow through the heterostructure, especially at low temperatures where thermally excited carriers are quenched. To ensure low leakage current and avoid Joule heating, we also fabricated heterostructures with an additional insulator layer, composed of FLG/hexagonal boron nitride (hBN)/CGT/FLG (Fig. S2). The resistance measured during cooling under a constant voltage of 1 V (Fig. 1c) from the FLG/CGT/FLG sample shows an increase in the resistance with decreasing temperature, characteristic of an insulator. As shown in the inset of Fig. 1c, a local minimum in the resistance gradient at $T$ ~ 55 K occurs close to the $T_C$, indicating a correlation between magnetic spin structures and electrical transport properties.

Figure 1d shows an over-focused LTEM image taken after 50 mT magnetic field cooling in the absence of an electric field. Hexagonally-close-packed skyrmionic bubbles with ~150 nm diameter were observed. When the exfoliated flakes are imaged along the *c*-axis, the bubble core (magnetization direction up) and peripheral (magnetization direction down) spin components do not generate any contrast in the LTEM imaging as they align with the imaging direction. Instead, the visible LTEM contrast arises from the Bloch walls, where the magnetic spins are perpendicular to the direction of imaging electrons, consistent with previous LTEM studies.[8,27] Based on the magnetic contrast in Fig. 1d, two distinguishable skyrmionic bubbles can be identified, as marked with the blue and red boxes. In an over-focus imaging condition, a CCW skyrmionic bubble displays a dark-to-bright contrast from the perimeter to the core due to the Lorentz-force-induced converging electron beams. Conversely, for a CW skyrmionic bubble, the contrast is reversed to bright-to-dark due to the diverging electron beams. Since the LTEM contrast also reverses when transitioning from underfocus to overfocus (Fig. S3), we used over-focus imaging to determine the chirality of skyrmionic bubbles. Here, all skyrmionic bubbles have their core spin direction antiparallel to the applied magnetic field, hence the polarity (P, the core spin orientation) of skyrmionic bubbles is always positive (+P), as schematically shown in Fig. 1e. Helicity is defined as the in-plane rotation sense, either CW or CCW, within the Bloch walls that separate the core domain from the peripheral domain (Fig. 1e). Note that the polarity in skyrmionic bubbles is fixed in our experiment. This was confirmed by micromagnetic simulations and LTEM contrast simulations, as shown in Fig. 1e. In our micromagnetic simulations, right-handed and left-handed skyrmionic bubbles are induced by negative and positive bulk DMIs ($\pm 1.25 \times 10^{-4}$ J/m$^2$), respectively. It should be noted that nonzero bulk DMI values are used to ensure controlled formation of CW and CCW bubbles for comparison with experimental LTEM contrast.

To study the effects of electric fields on skyrmionic bubble helicity, electric fields of different magnitude were applied during separate cooling processes from $T > 80$ K to 12 K through $T_C$ under a constant magnetic field of 50 mT. After the cooling process, the magnetic field was reduced to a residual value of approximately ~ 11 mT inside our electron microscope.[8] It should be noted that the directions of the electric and magnetic fields are parallel to the crystallographic *c*-axis and the imaging direction in the electron microscope (see Fig. 1b). We define the positive electric field direction as parallel to the imaging direction, and negative direction as antiparallel. Figure 2a-b show LTEM images of the FLG/CGT/FLG heterostructure taken at T = 14 K after cooling with a 50 mT magnetic field and simultaneous application of positive and negative 18 mV/nm electric fields, respectively. Based on the LTEM image contrast of each bubble, which exhibited either bright-to-dark or dark-to-bright transitions from the perimeter to the core, we found that the negative electric field induced CCW skyrmionic bubbles (Fig. 2a), while the positive electric field induced CW bubbles. This demonstrates that the helicity of skyrmionic bubbles has been selectively controlled by an external electric field throughout the field cooling process. As shown in Fig. 2c, attempts to reverse the helicity of the CW skyrmionic bubbles (Fig. 2b) by applying a negative electric field up to 36 mV/nm at 16 K, far below $T_C$, resulted in sample damage. The application of such a high electric field caused Joule heating, leading to a significant alteration in the magnetic domain structure characterized by coexisting stripe domains and skyrmionic bubbles (Fig. 2c), suggesting that the sample was heated above $T_C$. Importantly, the helicity of skyrmionic bubbles could be reversed from CW to CCW (see Fig. 2c) if the sample was heated and then cooled under an applied positive electric field (36 mV/nm) and 11 mT magnetic field. This underscores that the presence of nonzero DMIs is required during formation of skyrmionic bubbles for helicity selection.

To reduce current through the sample under high external electric fields, we fabricated a FLG/hBN/CGT/FLG heterostructure and performed LTEM imaging *in situ* during the field cooling processes (Supplementary Videos S1-S2 and Figs. S4-S5). Figs. 2d-h show LTEM images obtained after different field cooling processes, with electric fields ranging from -43 mV/nm to +36 mV/nm. The presence of the hBN layer allows the application of a broader range of electric fields without heating or damaging from leakage current with the voltage drop still mainly being across the CGT layer (see Supplementary Information for electric field calculation). A FLG/hBN/CGT/hBN/FLG heterostructure with an extra hBN may enable the application of a larger electric field before the sample breaks down. A substantial negative electric field of -43 mV/nm (Fig. 2d) predominantly induced CCW skyrmionic bubbles. Conversely, when a positive electric field of +36 mV/nm was applied, only CW skyrmionic bubbles were observed (Fig. 2h), confirming the helicity dependence on the electric field direction. With intermediate electric fields, both CCW and CW skyrmionic bubbles coexist after field cooling. In Fig. 2i, we quantify the percentage of CCW skyrmionic bubbles in a wide field of view (Fig. S6) as a function of electric field, finding almost a linear dependence on the electric field. We observe nearly random helicity (50% CCW and 50% CW) under an electric field of -18 mV/nm. This electric field offset may be attributed to an electron-beam-induced charging in the sample. In TEM, electron-transparent samples, especially if insulating, are often positively charged by secondary and/or Auger electron emission.[28] Thus, the 500 nm thick silicon nitride substrate is likely positively charged under electron beam illumination, resulting in a net positive electric field. This electron beam induced charging explains our previous study, which reported the observation of skyrmionic bubbles with single helicity observed in CGT flakes on silicon nitride membranes.[8]

The electric field-induced helicity of skyrmionic bubbles remained stable even after the electric field was fully turned off and magnetic field was reduced to 11 mT. Figures 2j-k show two LTEM images obtained before and after increasing the magnetic field from 11 mT to 113 mT, following a field cooling with a -18 mV/nm electric field and 50 mT magnetic field. Although the increased magnetic field caused the skyrmionic bubbles to shrink, their LTEM contrast remained unchanged. This indicates that the helicity and polarity were preserved despite variations in magnetic field.

To understand the microscopic origin of helicity selection, we conducted DFT calculations using SIESTA-TB2J[29], focusing on DMIs in CGT under external electric fields. The magnetic $Cr^{3+}$ ions in the edge-shared Te octahedra are arranged in a honeycomb lattice. Considering the honeycomb lattice as two interdigitated triangular lattices, there are two distinguishable Cr sites. In our DFT study, we identified two primary DMIs: nearest neighbor (NN) and next nearest neighbor (NNN). In the absence of an electric field, the NN interactions involving two $Te^{2-}$ anions do not exhibit a net DMI due to the centrosymmetric interaction pathway, as schematically shown in Fig. 3a. When an electric field is applied, the insulating CGT crystal becomes polarized by ionic displacements and/or electronic polarization. In a simple ionic charge screening scenario, the $Cr^{3+}$ cations move along the electric field direction while the $Te^{2-}$ anions shift in the opposite direction, resulting in the shortening of two Cr-Te bonds and the stretching of the other two. In Fig. 3a-c, the stretched Cr-Te bond lengths are indicated with gray lines. These ionic displacements break the inversion symmetry in the NN interaction pathway, resulting in nonzero DMIs. The primary components of electric-field-induced NN DMIs are in-plane and perpendicular to the nearest Cr-Cr bonds (see Table S1 in the Supplementary Information). In Figs. 3b and 3c, we show the calculated NN DMIs induced by negative and positive electric fields ($\pm 2 \times 10^4$ mV/nm) for two Cr sites in the honeycomb lattice. Similar electric-field-induced DMI vector patterns have been calculated for $CrI_3$ crystals, which also have a honeycomb lattice of Cr in edge-shared I octahedra.[17] A structural distinction between CGT and $CrI_3$ lies in the presence of the Ge dimer at the center of the Cr honeycomb lattice, which does not significantly influence exchange interactions. We found that the vector sums of in-plane components of NN DMIs are relatively small while the vector sum of out-of-plane components remains finite: +0.0737 meV and -0.0730 meV (Table S1) for negative and positive electric fields, respectively. Note that the resultant out-of-plane components of NN DMIs align with the applied electric field direction and consequently change their sign when the electric field direction is reversed, as shown in Fig. 3d. This vector sum of out-of-plane components of NN DMIs dictates the rotation sense of in-plane magnetic spins within the Bloch domain walls. Considering the DMI energy term $-D_{ij} \cdot (S_i \times S_j)$, CCW and CW spin arrangements are energetically favored for negative and positive electric fields, respectively, consistent with our experimental observations and micromagnetic simulations (Fig. 1e). According to our DFT calculations, the magnitude of the total NN DMIs scales linearly with the magnitude of electric field (see Fig. S7b). Consequently, the helicity response is linear with the magnitude of the external electric field, as demonstrated in Fig. 2i. The magnitude of electric fields applied in our DFT calculations exceeds those employed in the experiments by two orders of magnitude (Fig. 2). In experiments, not only is there a limitation in the maximum external voltage applied due to dielectric breakdown or leakage currents, but the external voltage applied may also not be entirely across the CGT layer due to air gaps, contact resistances, and other imperfections within the heterostructure. It also suggests that the resultant electric field-induced DMIs in experiments are not strong enough to alter the helicity of pre-existing skyrmionic bubbles, as observed in Figs. 2b-c. However, they may be potent enough to influence helicity selection at the onset of skyrmionic bubble formation during the field cooling process, as shown in Figs. 2d-h. Thus, although there is a discrepancy in electric field magnitude between theory and experiments, our study qualitatively demonstrates electric-field effects on helicity selection in skyrmionic bubbles. We rule out the effects of NNN DMIs on skyrmion helicity, as all NNN DMI vectors cancel each other regardless of the electric field direction (see Figs. 3b - c and Table S1).

Finally, we study skyrmionic bubble formation while field cooling the CGT across the $T_C$ with an applied electric field. In Fig. 4, we present LTEM images showing the skyrmionic bubble formation process during electric and magnetic field cooling. At ~ 54 K, slightly below $T_C$, small-diameter skyrmionic bubbles are prominent, as depicted in Fig. 4e (see also Supplementary Videos 1-2 and Figs. S4-S5). These small skyrmionic bubbles undergo dynamic formation and annihilation processes down to 35 K. For example, we observed CCW and CW skyrmionic bubbles merging into a CW bubble, as shown in Figs. 4b-c. The average radius gradually increases from ~ 67 nm to ~ 77 nm as the temperature decreases. The external magnetic field of 50 mT applied during the cooling process likely suppresses the formation of bubbles with large sizes, resulting in a narrow size distribution. Upon further cooling, we observed continuous fluctuations in the distance between neighboring skyrmionic bubbles down to 15 K, as shown in Fig. 4f. Overall, *in situ* LTEM imaging provides direct observation of the formation and evolution of topologically nontrivial skyrmionic bubbles during magnetic phase transitions.

In conclusion, we studied the effect of electric fields on the helicity of skyrmionic bubbles in 2D vdW CGT. Using electron microscopy imaging of skyrmionic bubble formation across the phase transition, we demonstrated that the helicity of these bubbles can be controlled by the direction of the electric field during magnetic field cooling. The resultant skyrmionic bubbles and their helicity remained stable at low temperatures, even with variations in applied magnetic and electric fields. The proportion of skyrmionic bubbles in CW versus CCW configurations exhibited almost a linear scaling with the electric field magnitude, facilitated by the insertion of a current-blocking hBN layer. According to our DFT calculations, the collective vector sum of NN DMIs aligns in the out-of-plane direction and reverses polarity when the direction of the electric field is altered, playing a pivotal role in chirality selection. Our study demonstrates that external electric fields, applied along the out-of-plane direction, can effectively control the in-plane spin helicity of nonmetallic 2D vdW magnets. This capability for electric field control of skyrmionic bubble chirality paves the way for developing advanced topological spintronic and quantum computing devices that utilize topological spin textures in 2D vdW magnets.

## Supporting Information

The Supporting Information is available.

Details for crystal growth, TEM sample and device fabrication, in situ cryogenic TEM experiments, density functional theory calculations, data analysis, micromagnetic and LTEM image simulations, two videos acquired during field cooling process, and estimates of electric fields in the FLG/hBN/CGT/FLG heterostructure.

## Reference:

1. Tokura, Y. and Kanazawa, N. Magnetic skyrmion materials. *Chemical Reviews* 121, 2857-2897 (2021).
2. Marrows, C.H. and Zeissler, K. Perspective on skyrmion spintronics. *Applied Physics Letters* 119, 250502 (2021).
3. Wang, Q.H. et al. The magnetic genome of two-dimensional van der Waals materials. *ACS Nano* 16, 6960-7079 (2022).
4. Fert, A., Reyren, N. and Cros, V. Magnetic skyrmions: advances in physics and potential applications. Nature Reviews Materials **2**, 17031 (2017).
5. Gibertini, M. et al. Magnetic 2D materials and heterostructures. Nature Nanotechnology 14, 408 (2019).
6. Avsar, A. et al. Spintronics in graphene and other two-dimensional materials. Reviews of Modern Physics 92, 021003 (2020).
7. Kurebayashi, H. et al. Magnetism, symmetry and spin transport in van der Waals layered systems. Nature Review Physics 4, 150-166 (2022).
8. Han, M.-G. et al. Topological magnetic-spin textures in two-dimensional van der Waals $Cr_2Ge_2Te_6$. Nano Letters 19, 7859-7865 (2019).
9. Park, T.-E. et al. Néel-type skyrmions and their current-induced motion in van der Waals ferromagnet-based heterostructures. Physical Review B **103**, 104410 (2021).
10. Wu, Y. et al. Néel-type skyrmions in $WTe_2/Fe_3GeTe_2$ van der Waals heterostructure. Nature Communications 11, 3860 (2020).
11. Nagaosa, N. and Tokura, Y. Topological properties and dynamics of magnetic skyrmions. Nature Nanotechnology 8 (12), 899-911 (2013).
12. Tokunaga, Y. et al. A new class of chiral materials hosting magnetic skyrmions beyond room temperature. Nature Communications 6, 7638 (2015).

13. Sierra, J. F. et al. Van der Waals heterostructures for spintronics and opto-spintronics. Nature Nanotechnology 16, 856-868 (2021).
14. Verzhbitskiy, I. and Eda, G. Electrostatic control of magnetism: Emergent opportunities with van der Waals materials. Applied Physics Letters 121, 060501 (2021).
15. Fillion, C.-E. et al. Gate-controlled skyrmion and domain wall chirality. Nature Communications 13, 5257 (2022).
16. Psaroudaki, C., Peraticos, E., and Panagopoulos, C., Skyrmion qubits: Challenges for future quantum computing applications, Appl. Phys. Lett. 123, 260501 (2023).
17. McGuire, M. A. et al. Coupling of crystal structure and magnetism in the layered, ferromagnetic insulator $CrI_3$. Chemistry of Materials 27, 612-620 (20156).
18. Liu, J. et al. Analysis of electric-field-dependent Dzyaloshinskii-Moriya interaction and magnetocrystalline anisotropy in a two-dimensional ferromagnetic monolayer. Physical Review B 97, 054416 (2018).
19. Behera, A. K., Chowdhury, S., and Das, S. R. Magnetic skyrmions in atomic thin $CrI_3$ monolayer. Applied Physics Letters 114, 232402 (2019).
20. Gong, C. et al. Discovery of intrinsic ferromagnetism in two-dimensional van der Waals crystals. Nature 546, 265-269 (2017).
21. Li, C.-K., Yao, X.-P., and Chen, G. Writing and deleting skyrmions with electric fields in a multiferroic heterostructure. Physical Review Research 3, L012026 (2021).
22. Martinez-Carracedo, G., et al. Tuning magnetic exchange interactions in two-dimensional magnets: The case of $CrGeX_3$ (X = Se, Te) and Janus $Cr_2Ge_2(Se,Te)_3$ monolayers. Physical Review B 110, 184406 (2024).
23. Verbitsky, I. A. et al. Controlling the magnetic anisotropy in $Cr_2Ge_2Te_6$ by electrostatic gating. Nature Electronics 3, 460-465 (2020).
24. Jiang, N. et al. Electric current control of spin helicity in an itinerant helimagnet. Nature Communications 30, 1601 (2020).
25. Fillion, C.-E., et al. Gate-controlled skyrmion and domain wall chirality. Nature Communications 13, 5257 (2022).
26. Dai, B., et al. Electric field manipulation of spin chirality and skyrmion dynamic. Science Advances 9, eade6836 (2023).
27. McCray, A. R. C., et al. Direct observation of magnetic bubble lattices and magnetoelastic effects in van der Waals $Cr_2Ge_2Te_6$. Advanced Functional Materials 33, 2214203 (2023).
28. Reimer, L. Image formation in low voltage scanning electron microscopy pp. 71 - 135. SPIE Press, Bellingham (1993).
29. He, X. et al., TB2J: a python package for computing magnetic interaction parameters. Computer Physics Communications, 107938 (2021).

**Acknowledgments**

The work at the Brookhaven National Laboratory was supported by the U.S. Department of Energy (DOE), Basic Energy Sciences, Materials Science and Engineering Division, under Contract DESC0012704. This research used Electron Microscopy resources of the Center for Functional Nanomaterials (CFN), which is a U.S. DOE Office of Science User Facility, at Brookhaven National Laboratory under Contract DESC0012704. The work at MIT was primarily supported through the Department of Energy BES QIS program on "Van der Waals Reprogrammable Quantum Simulator" under award number DE-SC0022277 and partially supported by the Quantum Science Center (QSC), a National Quantum Information Research Center of the U.S. Department of Energy (DOE) on probing quantum matter. P.N. gratefully acknowledges support from the John Simon Guggenheim Memorial Foundation (Guggenheim Fellowship) as well as support from a NSF CAREER Award under Grant No. NSF-ECCS-1944085. Z.S. was supported by ERC-CZ program (project LL2101) from Ministry of Education Youth and Sports (MEYS) and used large infrastructure from project reg. No. CZ.02.1.01/0.0/0.0/15_003/0000444 financed by the EFRR.

**Author contributions**

M.-G.H. and J.D.T. conceived the research. L.D. and J.S. synthesized and characterized the crystals. J.D.T. designed and fabricated heterostructure samples. F.C. contributed to electrical connection to the TEM samples and transport measurements of them. M.-G.H. and J.D.T., performed *in situ* TEM experiments. J.P.P., E.P. and P.N. carried out the DFT calculations. C.L. performed the micromagnetic and LTEM simulations. M.-G.H., J.D.T. and J.M. processed and analyzed data. F.M.R., P.N., and Y.Z. supervised the research. M.-G.H. wrote the paper with key contributions from J.D.T., E.P. F.M.R., and Y.Z. The manuscript reflects the contributions and ideas of all authors.

**Competing interests**

The authors declare no competing interests.

**Additional information**

**Correspondence and requests for materials** should be addressed to Myung-Geun Han.

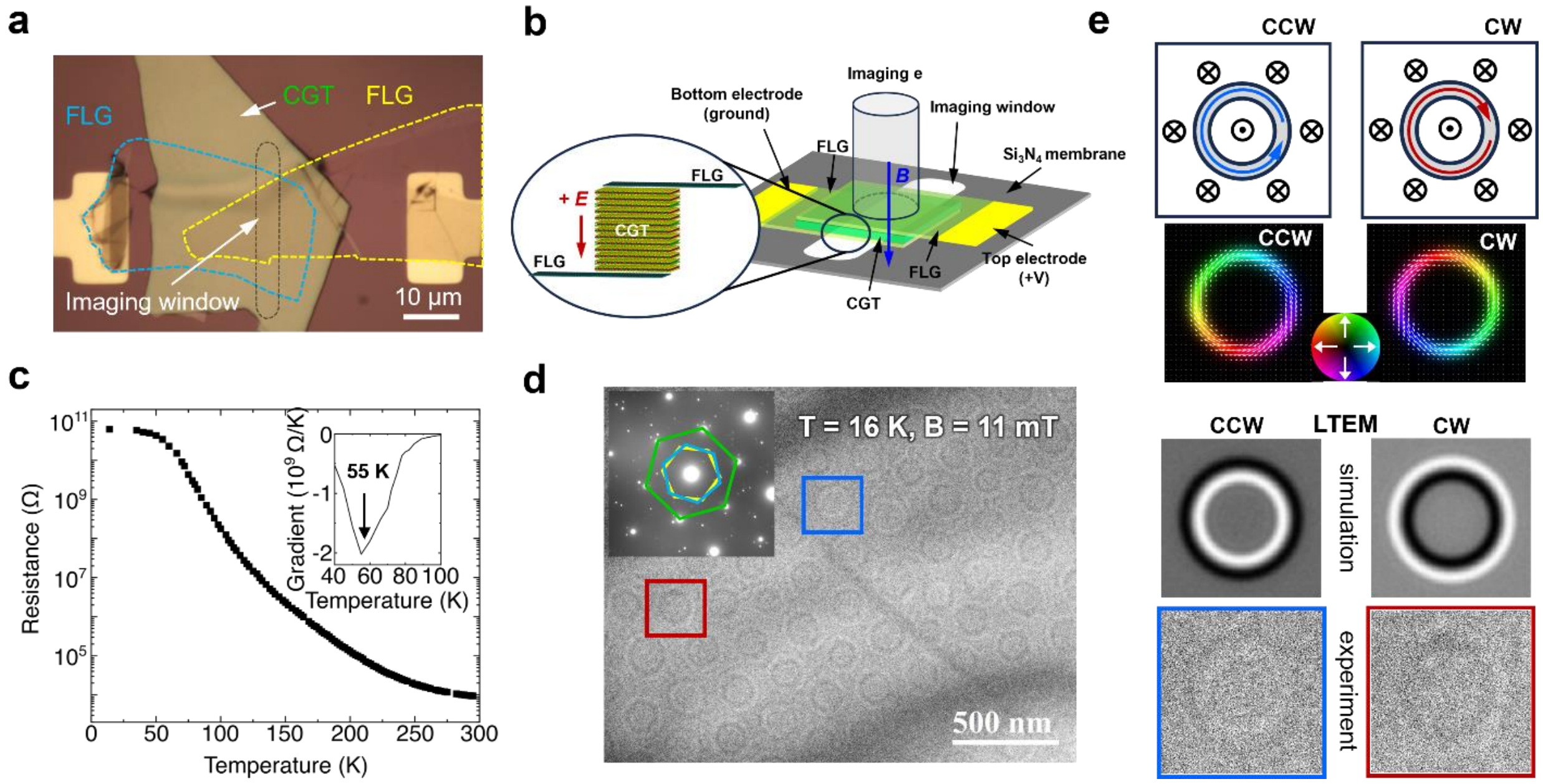

**Fig. 1| Heterostructure of FLG/$Cr_2Ge_2Te_6$ (CGT)/FLG for *in situ* biasing experiments and random helicity in skyrmionic bubbles. a,** An optical image of the FLG/CGT/FLG heterostructure on a custom-designed 500 nm thick silicon nitride membrane with a 4 μm x 34 μm observation window and two 50 nm thick Cr/Pt electrodes. Note that the CGT is sandwiched between two FLG layers contacted to the bottom and top Cr/Pt electrodes. **b,** A schematic of the device showing the direction of external electric and magnetic fields. The magnetic field in the imaging direction was controlled by the objective lens. **c,** Resistance vs. temperature data measured from the fabricated TEM sample. The inset shows the gradient of resistance as a function of temperature, which shows an upturn near 55 K. **d,** LTEM image showing skyrmionic bubble lattices obtained at T = 16 K and 11 mT magnetic field after the sample was cooled in a 50 mT field without an electric field. The defocus value was about +300 μm. The inset shows the electron diffraction pattern of the device. The diffraction spots from the two FLG layers (yellow and blue) and the CGT (green) are marked. **e,** Schematics of magnetic spin configuration of skyrmionic bubbles with right-handed and left-handed helicity and their in-plane spin configurations simulated by micromagnetic simulation (top panel). In order to induce the CCW bubble, a bulk DMI energy density of $-1.25 \times 10^{-4}$ J/m$^2$ was applied while the CW bubble was induced by $+1.25 \times 10^{-4}$ J/m$^2$ (the sign convention is the same as the electric field direction). The color wheel represents the direction and magnitude of in-plane spin components. Simulated LTEM image contrast and experimental LTEM images of skyrmionic bubbles with opposite helicity are shown in the bottom panel.

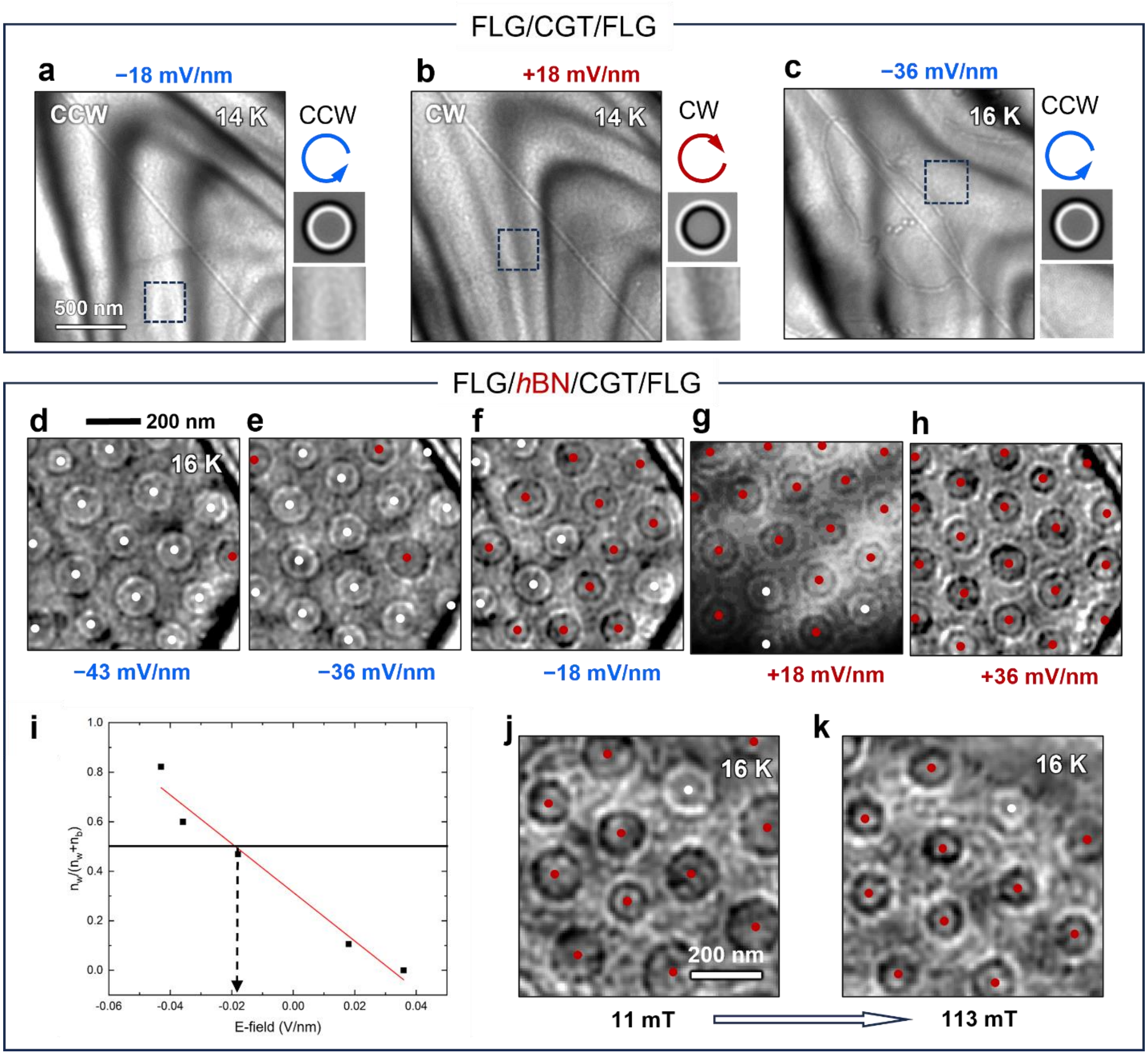

**Fig. 2| Electric field control of helicity in skyrmionic bubbles. a-b,** LTEM images (approx. 300 µm overfocus) of FLG/CGT/FLG heterostructure obtained after cooling in -/+ 18 mV/nm electric field and 50 mT magnetic field. **c,** LTEM image after poling the skyrmionic bubbles shown in **b** with large negative electric field (-36 mV/nm) at T = 16 K. Damaged parts of FLG layers can be seen. Domain structures completely changed from **b**, showing stripes and skyrmionic bubbles with reversed helicity (from CW in **b** to CCW in **c**) in the CGT layer. It should be noted that the LTEM images shown in **a** and **b** are obtained with the same + 300 µm overfocus condition. **d-h,** LTEM images (approx. 200 µm overfocus) of FLG/hBN/CGT/FLG heterostructure obtained after various electric field cooling. Here, the CW skyrmionic bubbles are marked with red dots while the CCW skyrmionic bubbles are marked with white dots. The skyrmionic bubble helicity induced by electric field is consistent with the results in **a** and **b**. **i**, Relative population of white and bright skyrmionic bubbles as a function of electric field during field cooling. **j-k**, Two LTEM images obtained after electric field 18 mV/nm under two different magnetic fields, 11 mT (left)

and 113 mT (right) showing no change in bubble contrast hence no change in polarity (core spin direction).

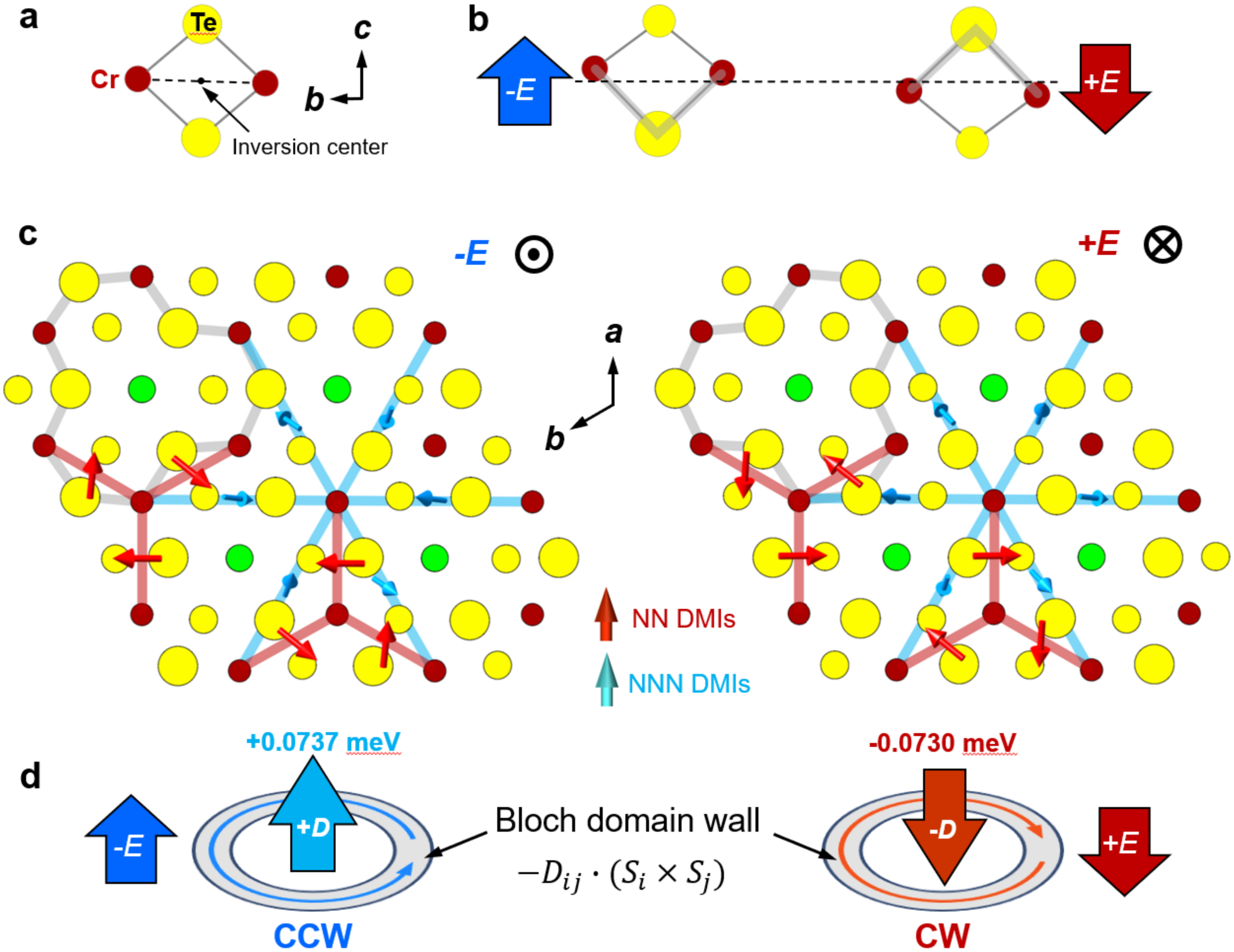

**Fig. 3| DMI vector patterns under external electric fields. a**, An atomic model showing the NN pathway under zero electric field. An inversion center is at the midpoint between two interacting Cr ions. **b**, Atomic models of the NN pathways under negative and positive electric fields, respectively. Ionic displacements are exaggerated for visualization. The stretched Cr-Te bonds are highlighted with gray lines. **c**, DMI patterns in the honeycomb lattice under negative (left) and positive (right) electric fields. The light blue arrows represent NNN DMIs along the NNN pathways highlighted with the light blue lines. The red arrows represent NN DMIs for two Cr sites with the NN pathways highlighted with the red lines. The gray lines highlight the stretched Cr-Te bonds due to ionic displacements. The Te ions in the stretched bonds are enlarged for visualization. **d**, Schematics of in-plane spin arrangements within the Bloch domain walls under negative (left) and positive (right) electric fields. The vector sums of out-of-plane components of the NN DMIs are indicated with the blue and red arrows.

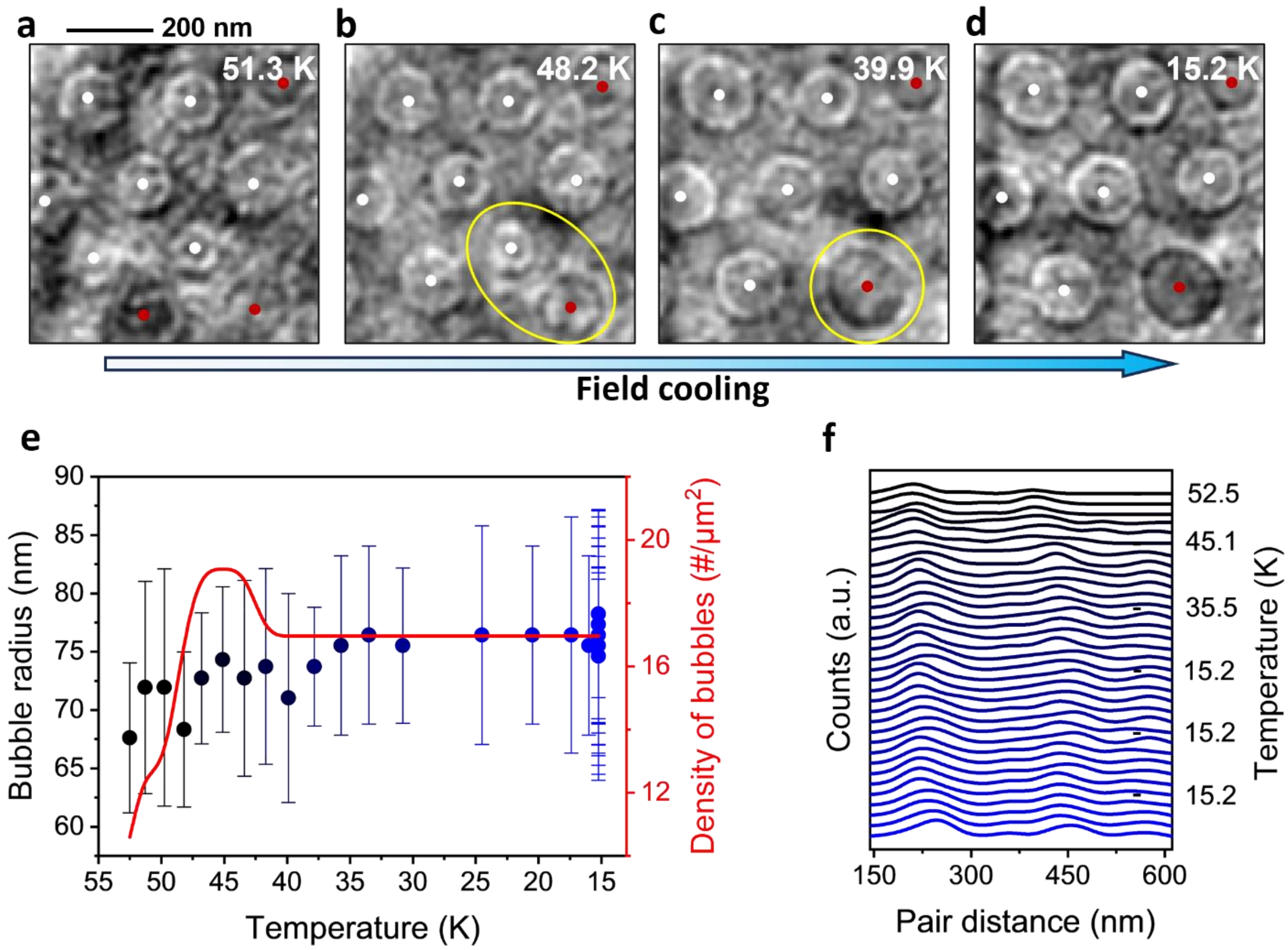

**Fig. 4| Helicity-selected bubble formation under field cooling. a-d,** A temperature series of LTEM images obtained during negative electric field (-36 mV/nm)) cooling. The magnetic field during cooling was 50 mT. For clarity, the CW skyrmionic bubbles are marked with red dots, while the CCW skyrmionic bubbles are marked with white dots. From 48.2 K (**b**) to 39.9 K (**c**), CW and CCW bubbles (circled in **b**) merge into a CCW bubble (circled in **c**). **e,** Bubble radius and bubble density as a function of temperature during field cooling. The data points were extracted from LTEM images. Small bubbles are first observed, and then average bubble size becomes uniform on further cooling down. The density of bubbles rapidly increases in the beginning before converging to ~ 17 bubbles/μm$^2$. **f,** The pair distance measurements across field cooling.

**TOC graphic**

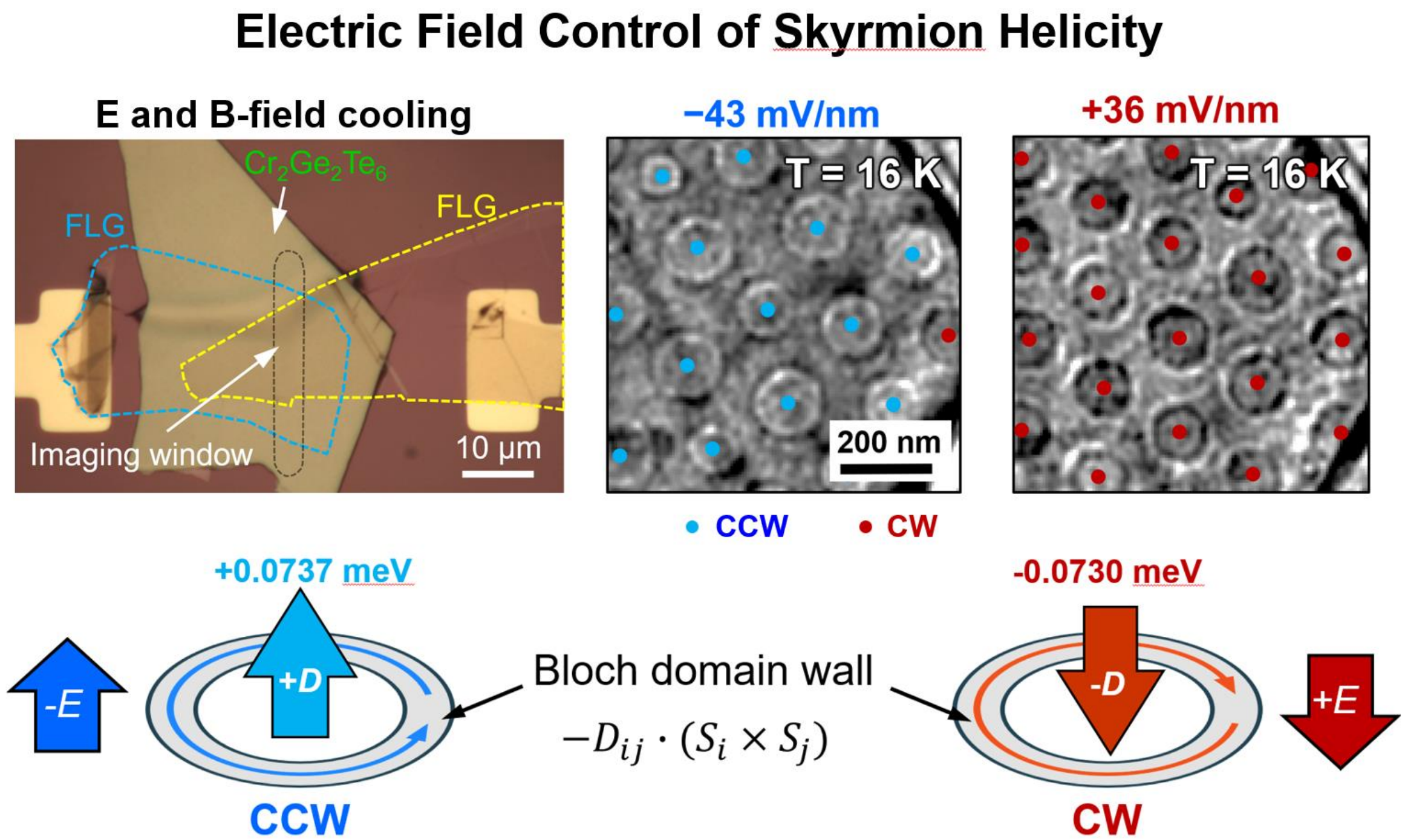

Electric Field Control of Skyrmion Helicity
E and B-field cooling
Cr2Ge2Te6
FLG
FLG
Imaging window
10 μm
−43 mV/nm
+36 mV/nm
T = 16 K
T = 16 K
200 nm
CCW
CW
+0.0737 meV
-0.0730 meV
-E
+D
Bloch domain wall
$-D_{ij} \cdot (S_i \times S_j)$
-D
+E
CCW
CW